\documentclass[aps,prb,twocolumn,floats,showpacs]{revtex4}

\usepackage{epsfig}
\usepackage{amsmath}
\usepackage{amssymb}

\newcommand{\bsigma}{\mbox{\boldmath$\sigma$}}

\newcommand{\be}{\begin{equation}}
\newcommand{\ee}{\end{equation}}
\newcommand{\bea}{\begin{eqnarray}}
\newcommand{\eea}{\end{eqnarray}}

\def\nn{\nonumber\\}

\def\rf#1{(\ref{#1})}

\def\vk{{{k}}}
\def\vp{{{p}}}
\def\vq{{{q}}}
\def\bsi{\boldsymbol\sigma}

\begin{document}

\title{A Luttinger Liquid Coupled to a Quantum Spin Bath: \\Flow Equation
  Approach to the Kondo Necklace Model} 

\author{F.H.L. Essler}
\affiliation{The Rudolf Peierls Centre for Theoretical Physics,
University of Oxford, 1 Keble Road, Oxford, OX1 3NP, United Kingdom}
\author{T. Kuzmenko}
\affiliation{The Rudolf Peierls Centre for Theoretical Physics,
University of Oxford, 1 Keble Road, Oxford, OX1 3NP, United Kingdom}
\author{I.~A.~Zaliznyak}
\affiliation{CMPMSD, Brookhaven National Laboratory, Upton, New York
11973-5000, USA.}

\begin{abstract}

We study a lattice realization of a Luttinger liquid interacting with a bath of
quantum spins in terms of an antiferromagnetic S=1/2 Heisenberg chain,
where each spin is also coupled to a $\sigma=1/2$ Kondo spin degree of freedom. 
This model describes the low-energy spin dynamics in quasi one dimensional
materials, where the electronic spins of the magnetic ions interact
with those of impurities, nuclei and possibly other spin species
present in their environment. For large ferromagnetic and
antiferromagnetic Kondo interaction $J'$ there are two phases
corresponding to an effective spin-1 Heisenberg chain and a dimerised 
spin-1/2 ladder, respectively. For weak Kondo interaction we establish
that the Kondo interaction drives the system to a strong coupling
regime. This suggests that $J'=0$ is the only critical point in the system.

\end{abstract}


\date{\today}
\maketitle

\section{Introduction}

Entanglement and cooperative spin behaviour induced by 
coupling two spin systems with different intrinsic dynamics is a
recurrent theme in condensed matter physics. In metals, it emerges in
the context of the coupling between the spins of itinerant 
electrons and localized atomic/impurity spins, known as Anderson
impurity or Kondo problem \cite{SchriefferWolff1966}. The famous
``central spin" problem of an electronic system interacting with a
bath of nuclear or impurity spins has recently re-emerged in the area of
spin-polarized microelectronics and the physics of quantum,
spin-entangled electronic states \cite{Childress_Science2006}.
Perhaps the most interesting example of this kind is a
quantum-critical spin system, such as found in strongly correlated
magnetic insulators, coupled to a bath of quantum or classical spins
\cite{Sachdev_Science1999,SachdevVojta_PRB2003,Lynn_PRB1990,Zheludev_PRL1998,Ronnow_Science2005,Zaliznyak_JETPL1996,Dumesh_JETP1999}.
In practice, the interaction between electronic spins in the system of
interest and ``external'' spin variables arises in a variety of
different contexts and occurs on vastly different energy scales.
The first example is the exchange interaction between valence
electrons involved in cohesion or chemical bonding and a localized
``d''-orbital, such as in the s-d model of magnetic metals.
If the d-orbital is singly occupied and the hybridization is weak, the
dominant interaction is Kondo exchange \cite{SchriefferWolff1966}.
Second is the coupling of electronic spins to impurity spins present
in real material. While its effects scale with impurity concentration,
in many cases and in particular at low temperatures it is the
determining mechanism for physical phenomena such as damping and
quantum decoherence. The response of a spin system that is either at
or close to quantum criticality to impurity spins is particularly
singular. It exhibits fascinating impurity-driven physics,
\cite{Sachdev_Science1999,SachdevVojta_PRB2003}, which has emerged
in studies of lightly doped cuprates and related two-dimensional Mott 
insulators. 
Another example occurs in complex alloys, where in addition to
magnetic $3d$ ions there often exist magnetic rare-earth cations 
(R$^{3+}$), which lead to a lattice of macroscopically many
``impurity" spins even for ideal stoichiometric materials
\cite{Lynn_PRB1990,Zheludev_PRL1998}. In the Haldane (S=1) chain
atiferromagnet R$_2$BaNiO$_5$, the cooperative coupling to
paramagnetic rare-earth spins dramatically modifies the spin 
dynamics of gapped Ni$^{2+}$ chains and induces magnetic order at a
finite temperature \cite{Zheludev_PRL1998}. 
Finally, at very low energies/temperatures the hyperfine coupling of 
electronic and nuclear spins becomes important
\cite{Ronnow_Science2005,Zaliznyak_JETPL1996,Dumesh_JETP1999}.
Indeed, many abundant isotopes of magnetic ions have non-zero
nuclear spin \cite{nuclear_spins}. Furthermore their electronic spins
can also interact with nuclei of surrounding ligand ions
\cite{AbrahamBleaney}.

Although in many cases the coupling of the spin system to external
spin degrees of freedom can be neglected in the same way as
the coupling to a generic thermostat is swept under the carpet in
equilibrium statistical mechanics, it is of the same fundamental
importance. The existence of such a coupling is required in order for
the quantum (spin) system to reach its equilibrium state, or the
ground state at T=0. Relaxation to equilibrium requires the change of
both energy and angular momentum, which are integrals of motion for an
isolated spin system. The simplest example is a Heisenberg magnet (or
paramagnet) in a magnetic field. Here the Hamiltonian conserves the
total spin component along the field direction. It is only through
(the implicit) coupling to some external system of angular momenta, or
spin bath, that the magnet can adjust its total spin as the magnetic
field changes (e.g., in a quantum phase transition from a spin-gap to
a magnetized phase).

While the effect of external spin degrees of freedom on the spin
dynamics can often be averaged out, e.g., in the framework of
spin-boson or spin-bath models
\cite{Legett_RMP1987,ProkofievStamp_RPP2000}, there are important
cases where the coupled dynamics of the two systems is 
of paramount interest. One such case is known as ``pulling" -- which
refers to the hybrid dynamics in a system of electronic spins coupled
to a thermalized ensemble of nuclear spins. Interest in this
phenomenon was recently renewed by studies of field-induced quantum
phase transitions in quantum magnets, where it was discovered that
pulling prevents the expected full softening of the electronic spin
excitation spectrum \cite{nucl_T}. Instead, the latter acquires a gap,
which increases with decreasing temperature, while a soft-mode
behaviour is induced in the nuclear spins, which "take over" the
quantum criticality 
\cite{Ronnow_Science2005,Zaliznyak_JETPL1996,Dumesh_JETP1999}.

\begin{figure}[ht]
\begin{center}
\noindent \epsfxsize=0.45\textwidth \epsfbox{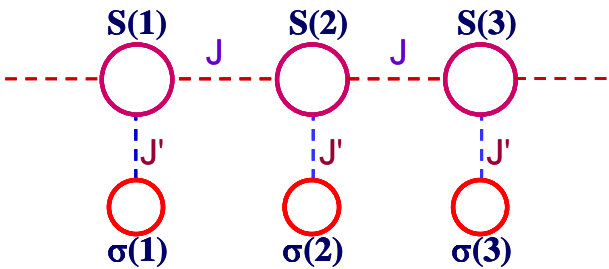}
\end{center}
\caption{One-dimensional Kondo necklace model: a ``master'' S = 1/2
Heisenberg chain with antiferromagnetic coupling $J$, where each spin
interacts with an extra, Kondo spin degree of freedom $\sigma = 1/2$,
via coupling $J'$.} \label{sp-ch}
\end{figure}

Here we focus on the opposite, much less studied and understood limit
of a quantum-critical spin system coupled to a bath of quantum spins,
whose dynamics is governed solely by their Berry phases. We consider
a Heisenberg S=1/2 chain with antiferromagnetic exchange coupling
$J>0$, where each spin is coupled to an additional local Kondo spin
$\sigma = 1/2$ by an exchange J$'$, 

\begin{equation}
H = J\sum_{\langle ij\rangle}^{N}{\bf S}(i) \cdot {\bf S}(j) +
J'\sum_{i=1}^{N}{\bf S}(i) \cdot \bsigma(i). \label{H0+H1}
\end{equation}
This model is sometimes referred to as incomplete ladder, or
SU(2) Kondo-necklace model, Fig. \ref{sp-ch}. The model \rf{H0+H1} is a
one-dimensional (1D) analog of the incomplete bilayer, which recently
received much attention in the context of the 2D cuprates
\cite{Hoglund_PRL2007}. The fundamental difference between the 1D and
2D Kondo necklaces is that in D=2 quantum criticality is achieved by
tuning the coupling J$'$, while in D=1 the S=1/2 chain is a critical
Luttinger liquid for J$'$ = 0. In what follows we allow the
Kondo-coupling $J'$ in (\ref{H0+H1}) to the local spins to be either
ferromagnetic or antiferromagnetic.

The model (\ref{H0+H1}) is closely related to the Kondo necklace model,
\begin{equation}
H_{\rm KNM} = J \sum_{\langle ij\rangle}^{N} S^x(i) S^x(j) + S^y(i)
S^y(j) + J'\sum_{i=1}^{N}{\bf S}(i) \cdot \bsigma(i), \label{HKNM}
\end{equation}
which was introduced in Ref. \onlinecite{Doniach} as a simplified
version of the Kondo lattice model in the 1D case.
The Kondo necklace model \rf{HKNM} has been studied by a variety of
methods such as Monte Carlo simulations \cite{Scal}, real-space
renormalization group (RG) techniques \cite{RG}, exact numerical
diagonalization \cite{Santini}, density-matrix renormalization group
(DMRG) computations \cite{DMRG,DMRG2}, bond operator mean field theory
\cite{Zhang,Langari} and bosonization \cite{Strong,Kiselev}. While all
methods agree in the regime $|J'|\gg J$, it is still controversial whether
a small
Kondo coupling $|J'|\ll J$ leads to the formation of a spin gap.
While real-space RG and exact diagonalization of small systems
suggest the existence of a critical value $J'_c$ of the Kondo
coupling strength, below which the spin gap vanishes, Monte-Carlo and
DMRG computations indicate that in fact $J'_c=0$. The doped case
has recently been investigated in \cite{Aristov}.
In the two and three-dimensional cases it is well established that
there is a quantum phase transition between a N\'eel ordered ground
state at small $J'>0$ and a Kondo-singlet phase at $J'>J'_c$
\cite{Zhang,Langari,Brenig,Matsushita,Kotov}. 

In the present work we study the model (\ref{H0+H1}) using the flow equation method.
In order to ``close'' the system of flow equations we employ a
decoupling scheme for spin operator products, which is asymptotically
valid in the limit $J' \rightarrow 0$, and use essentially exact expressions for
the two-spin correlation function in the ``master" spin-1/2 chain,
derived by field theory methods \cite{affleck98,LT}. The flow
equations obtained upon decoupling are then solved numerically in the weak
coupling limit. 

\section{Strong Coupling Limits}
In order to understand general properties and phase diagram of the
model, it is instructive to analyze the strong coupling limits
$J'\to\pm\infty$ first.

\subsection{Ferromagnetic Kondo coupling $J'\to-\infty$}
For strong ferromagnetic Kondo coupling an effective spin 1 is formed
on each rung. The Hamiltonian describing the interaction between these
spins takes the form of an antiferromagnetic spin-1 Heisenberg chain
\be
H_{\rm eff}=\frac{J}{4}\sum_j \vec{T}_j\cdot\vec{T}_{j+1}.
\ee
Here $T_j^\alpha$ are spin-1 operators. The antiferromagnetic
spin-1 chain is known to display a spontaneously broken $Z_2\otimes
Z_2$ symmetry characterized by a non-zero string order ${\cal   O}_{\rm
  string}^\alpha\neq 0$, where \cite{stringorder}
\be
{\cal O}^\alpha_{\rm string}=\lim_{n\to\infty}\left\langle
T^\alpha_j\exp\Bigl(i\pi\sum_{k=j+1}^{j+n-1}T^\alpha_k\Bigr)
T^\alpha_{j+n}\right \rangle.
\ee
It is well known that excitations in the spin-1 Heisenberg chain are
desribed in terms of a triplet of gapped magnons.
\subsection{Antiferromagnetic Kondo coupling $J'\to\infty$}
In this limit the ground state is that of decoupled singlet
dimers. This is in fact the same ground state as for the regular
two-leg ladder with $J_{\rm rung}\gg J_{\rm leg}$ in the limit
$J_{\rm rung}\to\infty$. It also can be characterized by a string
order parameter, e.g., \cite{GNT}
\bea
{\cal O}_{\rm string}&=&\lim_{n\to\infty}\left\langle
\prod_{k=j}^{j+n}\left[-4\sigma^z_jS^z_j\right]\right\rangle\nn
&=&\lim_{n\to\infty}\left\langle
\exp\Bigl(i\pi\sum_{k=j}^{j+n}\sigma^z_k+S^z_k\Bigr)\right
\rangle.
\eea
The excitation spectrum in this limit again has a gap.
\section{Weak Kondo Coupling}
The question we want to address is what happens for weak Kondo
couplings $|J'|\ll J$. The Kondo necklace model can be viewed as a
particular limit of an asymmetric two-leg ladder model, in which the
coupling along the first leg $J$ is much larger than the rung coupling
$J'$, which in turn is large compared to the exchange $J_2$ along the
second leg
\be
J\gg J'\gg J_2.
\ee
This case is difficult to analyze for the following reason.
Bosonizing the spin chains making up the two legs of the ladder
results in a two-flavour Luttinger liquid. However, the cutoff of this
theory is equal to $J_2$. The rung coupling $J'$ can then not be
treated as a perturbation of the two-flavour Luttinger liquid as it is
much larger than the cutoff of the latter. In the Kondo necklace model
the role of $J_2$ is played by the RKKY interaction induced by $J'$,
which is of order $J'^2/J\ll |J'|$.

In order to analyze the small $J'$ regime we have employed Wegner's
flow equation method \cite{Wegner,Glaz-Wil,Kehrein,Stein,Raas,Sommer}.

\subsection{Flow Equation Method}
In the flow equation method \cite{Wegner,Glaz-Wil} a one-parameter
family of
unitarily equivalent Hamiltonians $H(l)$ is constructed via the
differential equation
\begin{eqnarray}
\frac{dH(l)}{dl} &=& [\eta(l),H(l)].
\end{eqnarray}
The anti-hermitian generator $\eta(l)$ is taken as
\be
\eta(l)=[H_0(l),H(l)],
\ee
where $H_0(l)$ is a particularly chosen ``diagonal'' part of the
Hamiltonian. For the Kondo necklace model we chose $H_0(l)$ as

\begin{eqnarray}
H_0(l)&=&\frac{1}{N}\sum_{k}J_k(l)\ {\bf S}_{k}\cdot {\bf
S}_{-k}\nn
&+&\frac{1}{N}\sum_{k}{\alpha}_{k}(l)\ {\boldsymbol
{\sigma}}_{k}\cdot {\boldsymbol{\sigma}}_{-k}.
\end{eqnarray}
Here the Fourier transformed spin operators are defined as
\be
{\bf S}_k=\sum_n{\bf S}(n)e^{-ikn},\ \
{\bf S}(n)=\frac{1}{N}\sum_k{\bf S}_ke^{ikn}.\label{F-tr}
\ee

\begin{figure}[ht]
\begin{center}
\noindent
\epsfxsize=0.45\textwidth
\epsfbox{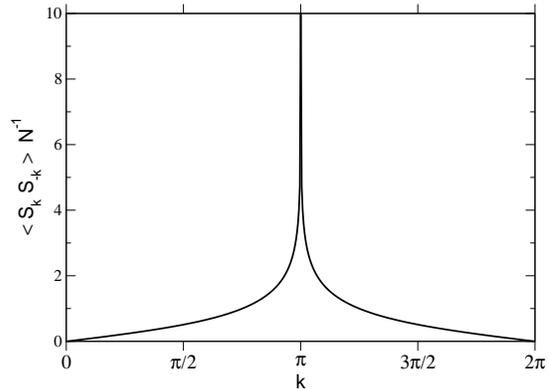}
\end{center}
\caption{\label{fig:strucfac}
Static spin-spin correlation function on the isotropic spin-1/2
Heisenberg chain.
}
\end{figure}

In the initial Hamiltonian \rf{H0+H1} the second term is absent, but it
will be generated under the flow. The full Hamiltonian will be of the
form
\be
H(l)=H_0(l)+H_1(l)+H_2(l)\ ,
\ee
where
\be
H_1(l)=\frac{1}{N}\sum_{k}J'_k(l){\bf S}_{k}\cdot {\boldsymbol
{\sigma}}_{-k}\ .
\ee
The contribution $H_2(l)$ will loosely speaking contain all
multi-spin interaction terms compatible with the global SU(2) spin
rotational symmetry. A key element in implementing the flow equation
approach is that the initial coupling $J'$ is small and all terms in
$H_2(l)$ will be of order $J'^2$ or higher. As long as we constrain
our attention to the small $J'$ limit, we may therefore neglect
$H_2(l)$ when calculating the generator $\eta(l)$ of the unitary
transformation
\be
\eta(l)=[H_0(l),H_1(l)]=\eta_1(l)+\eta_2(l).
\label{eta}
\ee
The explicit forms of $\eta_{1,2}(l)$ are
\bea
\eta_1(l)&=&\frac{i}{N^2}\sum_{kk'}J'_{k'}(l)[J_k(l)-J_{k+k'}(l)]\nn
&&\quad\qquad [{\bf S}_{k+k'}\times {\bf S}_{-k}]
\cdot \boldsymbol{\sigma}_{-k'}, \label{eta-j}\nn
\eta_2(l)&=&
\frac{i}{N^2}\sum_{kk'}J'_{k'}(l)[{\alpha}_k(l)-{\alpha}_{k+k'}(l)]\nn
&&\quad\qquad [{\boldsymbol{\sigma}}_{k+k'}\times
{\boldsymbol \sigma}_{-k}]\cdot {\bf S}_{-k'}. \label{eta-al}
\eea
Working out the required commutators of $\eta_{1,2}(l)$ with
$H_{0,1}(l)$ (see Appendix \ref{app:comm}) we arrive at the following
expression for the Hamiltonian
$H(l)$
\begin{eqnarray}
H(l)&=&\frac{1}{N}\sum_{k}J_k(l){\bf S}_{k}\cdot {\bf S}_{-k}
+\frac{1}{N}\sum_{k}\alpha_k(l)
{\boldsymbol{\sigma}}_{k}\cdot {\boldsymbol{\sigma}}_{-k}\nonumber\\
&+&\frac{1}{N}\sum_{k}J'_k(l){\bf S}_{k}\cdot {\boldsymbol
{\sigma}}_{-k}\nn
&+&\frac{1}{N^3}\sum_{\vk,\vp,\vq}
M^{(1)}_{k,p,q}(l)\
{\bf S}_{\vk+\vp+\vq}\cdot{\boldsymbol\sigma}_{-\vp}\
{\bf S}_{-\vk}\cdot{\boldsymbol\sigma}_{-\vq}\nn
&+&\frac{1}{N^3}\sum_{\vk,\vp,\vq}
M^{(2)}_{k,p,q}(l)\
{\bf S}_{\vk+\vp}\cdot{\bf S}_{\vq}\
{\bf S}_{-\vk}\cdot{\boldsymbol\sigma}_{-\vp-\vq}\nn
&+&\frac{1}{N^3}\sum_{\vk,\vp,\vq}
M^{(3)}_{k,p,q}(l)\
{\bf S}_{\vk+\vp+\vq}\cdot{\bf S}_{-\vk}\
{\boldsymbol\sigma}_{-\vq}\cdot{\boldsymbol\sigma}_{-\vp}\nn
&+&\frac{1}{N^3}\sum_{\vk,\vp,\vq}
M^{(4)}_{k,p,q}(l)\
{\bsi}_{\vk+\vp}\cdot{\bsi}_{\vq}\
{\bsi}_{-\vk}\cdot{\bf S}_{-\vp-\vq}\nn
&+&\frac{i}{N^3}\sum_{\vk,\vp,\vq}
M^{(5)}_{k,p,q}(l)\
\left[{\bf S}_{\vk+\vp+\vq}\times{\bf S}_{-\vk}\right]
\cdot{\boldsymbol\sigma}_{-\vp-\vq}\nn
&+&\frac{i}{N^3}\sum_{\vk,\vp,\vq}
M^{(6)}_{k,p,q}(l)\
\left[{\bsi}_{\vk+\vp+\vq}\times{\bsi}_{-\vk}\right]
\cdot{\bf S}_{-\vp-\vq}.\nn
\label{Hkl}
\end{eqnarray}
The initial values of the various couplings are
\begin{eqnarray*}
J_{k}(0)&=&J\cos{k},  \quad J'_{k}(0)=J',\quad
M^{(a)}_{k,p,q}(0)=0.
\end{eqnarray*}
In order to obtain a set of flow equations we need to decouple the
three- and four-spin terms. We do this by expanding in fluctuations
around the $J'=0$ ground state, i.e.,
\begin{eqnarray}
{\bf S}_{k}\cdot {\bf S}_{k'}\ {\boldsymbol {\sigma}}_{q}\cdot
{\boldsymbol {\sigma}}_{-k-k'-q}&=& \langle{\bf S}_{k}\cdot
{\bf S}_{k'}\rangle {\boldsymbol {\sigma}}_{q}\cdot {\boldsymbol
{\sigma}}_{-k-k'-q}\nn
&&\hskip-30pt+\langle {\boldsymbol{\sigma}}_{q}\cdot
{\boldsymbol
{\sigma}}_{-k-k'-q}\rangle {\bf S}_{k}\cdot {\bf S}_{k'}\nn
&&\hskip-30pt+:{\bf S}_{k}\cdot {\bf S}_{k'}\ {\boldsymbol {\sigma}}_{q}\cdot
{\boldsymbol {\sigma}}_{-k-k'-q}:\ ,
\eea
\bea
{\bf S}_{k'}\cdot {\boldsymbol {\sigma}}_{q}\ {\bf S}_{k}\cdot
{\boldsymbol {\sigma}}_{-k-k'-q}&=&
\frac{1}{3}\langle{\bf S}_{k'}\cdot {\bf S}_k\rangle {\boldsymbol
{\sigma}}_q\cdot{\boldsymbol \sigma}_{-k-k'-q}\nn
&&\hskip-30pt+\frac{1}{3}\langle {\boldsymbol{\sigma}}_{q}\cdot
{\boldsymbol
{\sigma}}_{-k-k'-q}\rangle{\bf S}_{k'}\cdot {\bf S}_{k}\nn
&&\hskip-30pt+:{\bf S}_{k'}\cdot {\boldsymbol {\sigma}}_{q}\ {\bf S}_{k}\cdot
{\boldsymbol {\sigma}}_{-k-k'-q}:\ ,
\eea
\bea
{\bf S}_{k}\cdot {\bf S}_{k'}\ {\bf S}_{q}\cdot {\boldsymbol
{\sigma}}_{-k-k'-q}&=& \langle {\bf S}_{k}\cdot {\bf
S}_{k'}\rangle{\bf S}_{q}\cdot {\boldsymbol {\sigma}}_{-k-k'-q}\nn
&&\hskip-30pt+ \frac{1}{3}\langle {\bf S}_{k}\cdot {\bf
S}_{q}\rangle{\bf S}_{k'}\cdot {\boldsymbol {\sigma}}_{-k-k'-q}\nn
&&\hskip-30pt+ \frac{1}{3}\langle {\bf S}_{k'}\cdot {\bf
S}_{q}\rangle{\bf S}_{k}\cdot {\boldsymbol {\sigma}}_{-k-k'-q}\nn
&&\hskip-30pt+:{\bf S}_{k}\cdot {\bf S}_{k'}\ {\bf S}_{q}\cdot {\boldsymbol
{\sigma}}_{-k-k'-q}:\ .
\label{fluct}
\end{eqnarray}
These expansions can be motivated by bosonizing the Hamiltonian
(\ref{Hkl}) (see Appendix \ref{app:boson}). Substituting \rf{fluct}
into \rf{Hkl} we obtain expressions for $H_0(l)$ and $H_1(l)$.
The static spin-spin correlation functions entering \rf{fluct} and
hence the expressions for $H_{0,1}(l)$ should be calculated
self-consistently with respect to the flowing Hamiltonian
$H_0(l)$. However, at weak coupling $|J'|\ll J$ one may calculate the
correlators with respect to the initial Hamiltonian $H_0(0)$, as the
corrections are of higher order in $J'$. Taking into account that
\bea
\langle{\bf S}_{k}\cdot {\bf
  S}_{k'}\rangle&=&\delta_{k,-k'}\langle{\bf S}_{k}\cdot {\bf
S}_{-k}\rangle\ ,\nn
\langle{\bsi}_{k}\cdot {\bsi}_{k'}\rangle&=&\frac{3N}{4}\delta_{k,-k'}\ ,
\eea
and retaining only terms quadratic in spin operators,
the Hamiltonian (\ref{Hkl}) is reduced to
\begin{eqnarray}
\tilde{H}(l)&=&\frac{1}{N}\sum_{k}J_k(l)\ {\bf S}_{k}\cdot {\bf
S}_{-k}+\frac{1}{N}\sum_{k}{\alpha}_{k}(l)\ {\boldsymbol
{\sigma}}_{k}\cdot {\boldsymbol{\sigma}}_{-k}\nn
&&+\frac{1}{N}\sum_{k}J'_k(l)\ {\bf S}_{k}\cdot
{\boldsymbol {\sigma}}_{-k}. \label{Hkl-simpl}
\end{eqnarray}
The flow equations for the couplings take the form
\begin{widetext}
\begin{eqnarray}
&&\frac{dJ_k}{dl}=
 \frac{2}{3 N}\sum_{k'}(J'_{k'})^2 (2J_k-J_{k+k'}-J_{k-k'})\frac{\langle
{\boldsymbol \sigma}_{k'}\cdot {\boldsymbol
\sigma}_{-k'}\rangle}{N}
+  \frac{2(J'_{k})^2}{3
N}\sum_{k'}(2\alpha_{k'}-\alpha_{k+k'}-\alpha_{k-k'})\frac{\langle
{\boldsymbol \sigma}_{k'}\cdot {\boldsymbol
\sigma}_{-k'}\rangle}{N},
\label{flow1}
\eea
\bea
&&\frac{d\alpha_k}{dl} =
 \frac{2}{3 N}\sum_{k'}(J'_{k'})^2
(2\alpha_k-\alpha_{k+k'}-\alpha_{k-k'})\frac{\langle
{\bf S}_{k'}\cdot {\bf S}_{-k'}\rangle}{N}
+
 \frac{2(J'_{k})^2}{3 N}\sum_{k'}(2J_{k'}-J_{k+k'}-J_{k-k'})\frac{\langle
{\bf S}_{k'}\cdot {\bf S}_{-k'}\rangle}{N},
\label{flow2}
\eea
\bea
&&\frac{dJ'_k}{dl}=
\frac{4J'_{k}}{3
N}\sum_{k'}\Big\{(J_{k'}-J_k)(J_{k+k'}+J_{k-k'}-2J_{k'})\frac{\langle
{\bf S}_{k'}\cdot {\bf S}_{-k'}\rangle}{N}
+(\alpha_{k'}-\alpha_k)
(\alpha_{k+k'}+\alpha_{k-k'}-2\alpha_{k'})\frac{\langle
{\boldsymbol \sigma}_{k'}\cdot {\boldsymbol
\sigma}_{-k'}\rangle}{N}\Big\}\nn
&&\ \ \ \ \ \ \; +\frac{4}{3
N}\sum_{k'}J'_{k'}J'_{k+k'}\Big\{(J_{k+k'}-J_{k})\frac{\langle
{\bf S}_{k+k'}\cdot {\bf S}_{-k-k'}\rangle}{N}
+ (\alpha_{k+k'}-\alpha_{k})\frac{\langle {\boldsymbol
\sigma}_{k+k'}\cdot {\boldsymbol
\sigma}_{-k-k'}\rangle}{N}\Big\}.
\label{flow3}
\end{eqnarray}
\end{widetext}
As we have remarked earlier, in order to solve the flow equations we
need to know the static spin-spin correlator $\langle{\bf S}_{k}\cdot
{\bf   S}_{-k}\rangle$ for the isotropic spin-1/2 Heisenberg chain.
This can be calculated accurately from the results of
Refs. [\onlinecite{affleck98,LT}]. There the large distance
asymptotics of spin-spin correlations functions was determined by
combining exact results for correlation amplitudes \cite{lukyanov}
with renormalization group improved perturbation theory in the
marginally irrelevant interaction of spin currents present in the
continuum description of the spin-1/2 Heisenberg chain that
\be
\langle {\bf S}(m+j)\cdot {\bf
  S}(j)\rangle\approx \frac{3}{4}
\left[\frac{(-1)^m}{m}\ \sqrt{\frac{2}{\pi^3  g}} f_1(g)
-\frac{f_2(g)}{\pi^2m^2}\right],
\label{spinspin}
\ee
where
\bea
f_1(g)&=& 1+\Big(\, \frac{3}{8}-\frac{c}{2}\,\Big)\, g
          +\Big(\,\frac{5}{128}-\frac{c}{16}
          -\frac{c^2}{8}\,\Big)\, g^2\nonumber\\
   &+&\Big(\,\frac{21}{1024}+\frac{7 c}{256}
          -\frac{7 c^2}{64}-\frac{c^3}{16}
          +\frac{13\, \zeta(3)}{32}\,\Big)\, g^3,\nn
f_2(g)&=&1+\frac{g}{2}+\Big(c+\frac{3}{4}\,\Big)\,
\frac{g^2}{2}+\frac{c(c+2)}{2}\, g^3.
\eea
Here $g$ is the running coupling constant
\be
\sqrt{g}\ e^{1/g}=2\sqrt{2\pi}\ e^{\gamma_E+c}\ m\ ,
\ee
and $c$ is a free parameter that related to the choice of
renormalization scheme. It was demonstrated in Ref. [\onlinecite{LT}]
that for $l\geq 2$ (\ref{spinspin}) is in very good agreement with
numerical results. Supplementing these results with the known values
\bea
\langle {\bf S}(j)\cdot {\bf  S}(j+1)\rangle&=&
\frac{1}{4}+\sum_{n=1}^\infty\frac{(-1)^n}{n}=-0.443147...\ ,\nn
\langle {\bf S}(j)\cdot {\bf  S}(j)\rangle&=&\frac{3}{4}\ ,
\eea
and then Fourier transforming, we arrive at the result shown in
Fig.\ref{fig:strucfac}.
We see that in momentum space the spin-spin correlator is dominated by
the logarithmic divergence at $k=\pi$.

\subsection{Solution of the Flow Equations}

We are now in a position to solve the flow equations
\rf{flow1}-\rf{flow3} numerically. We find that for small initial
values on the Kondo interaction $|J'|<0.1J$ the behaviour of the solutions to
the flow equations exhibits two regimes. In the following we measure
$l$ in units of $J^{-2}$. For large flow parameters $l\agt 100$
the couplings appear to approach a fixed point. However, as $l$
increases further, the behaviour eventually changes and the couplings
diverge. We interpret the eventual runaway flow as being indicative of
the emergence of a strong coupling phase, characterized by the
formation of a spectral gap. 

\subsubsection{Kondo Interaction}
\begin{figure}[ht]
\begin{center}
\noindent
\epsfxsize=0.45\textwidth
\epsfbox{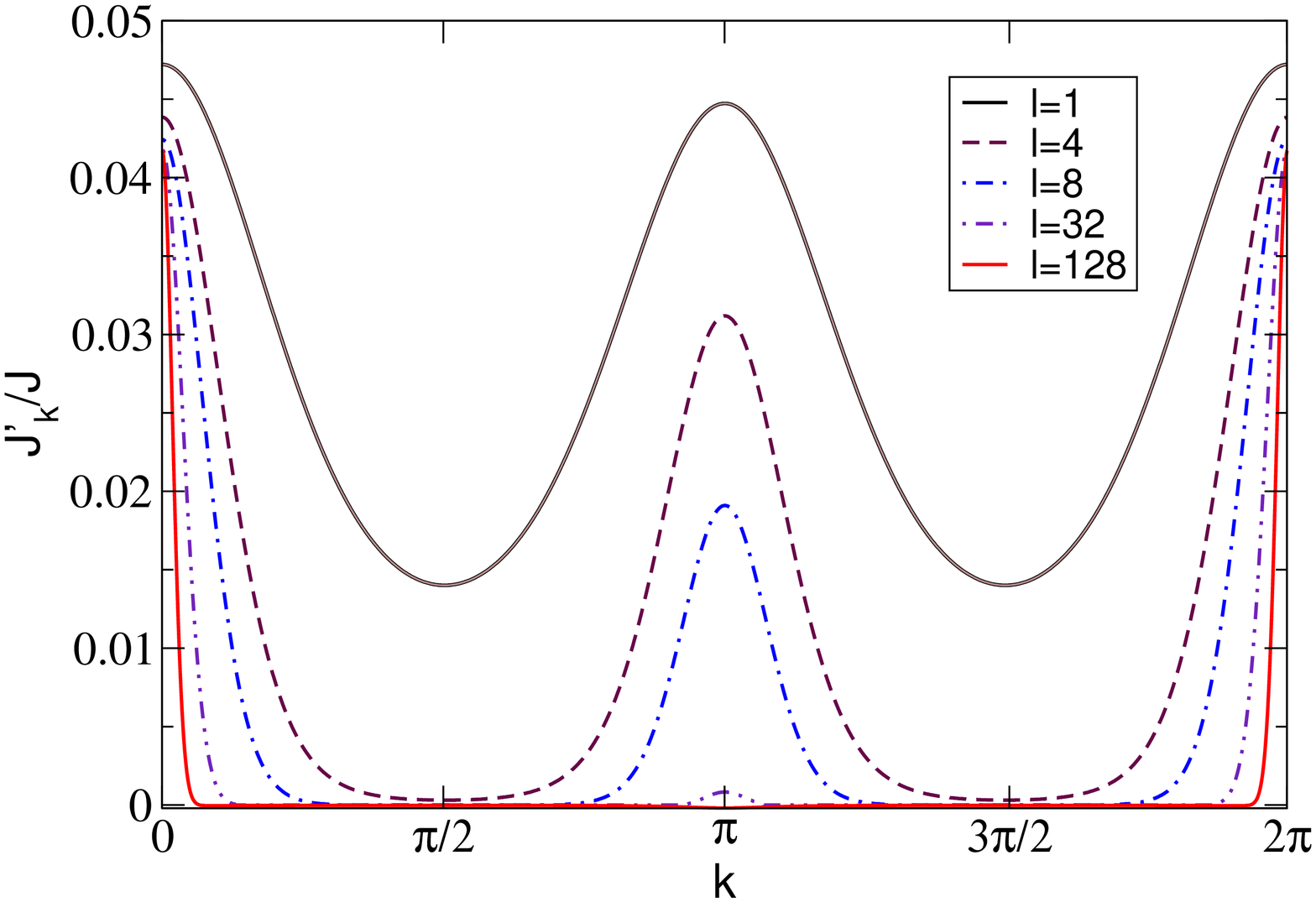}
\end{center}
\caption{Kondo coupling $J'_k$ as a function of the momentum $k$ for
several values of the parameter $l$ characterizing the flow
$l=1$, $l=4$, $l=8$, $l=32$ and $l=128$.
The initial value is $J'_k=0.05J$.  The Kondo interaction becomes small
under the flow except in the vicinity of $k=0,2\pi$.} 
\label{Jp-k}
\end{figure}

\begin{figure}[ht]
\begin{center}
\noindent
\epsfxsize=0.45\textwidth
\epsfbox{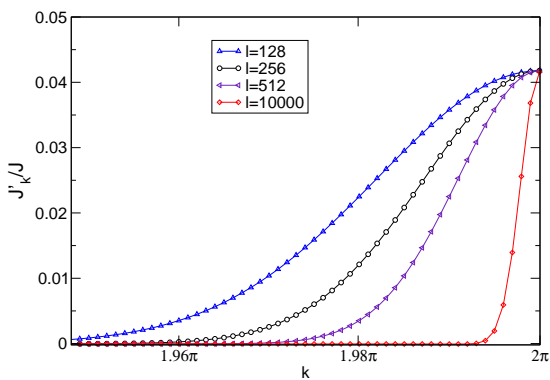}
\end{center}
\caption{Kondo coupling $J'_k$ in the vicinity of $k=2\pi$ for
several values of the parameter $l$ characterizing the flow. The
initial value is $J'_k=0.05J$.}
\label{Jp-k2pi}
\end{figure}

\begin{figure}[ht]
\begin{center}
\noindent
\epsfxsize=0.45\textwidth
\epsfbox{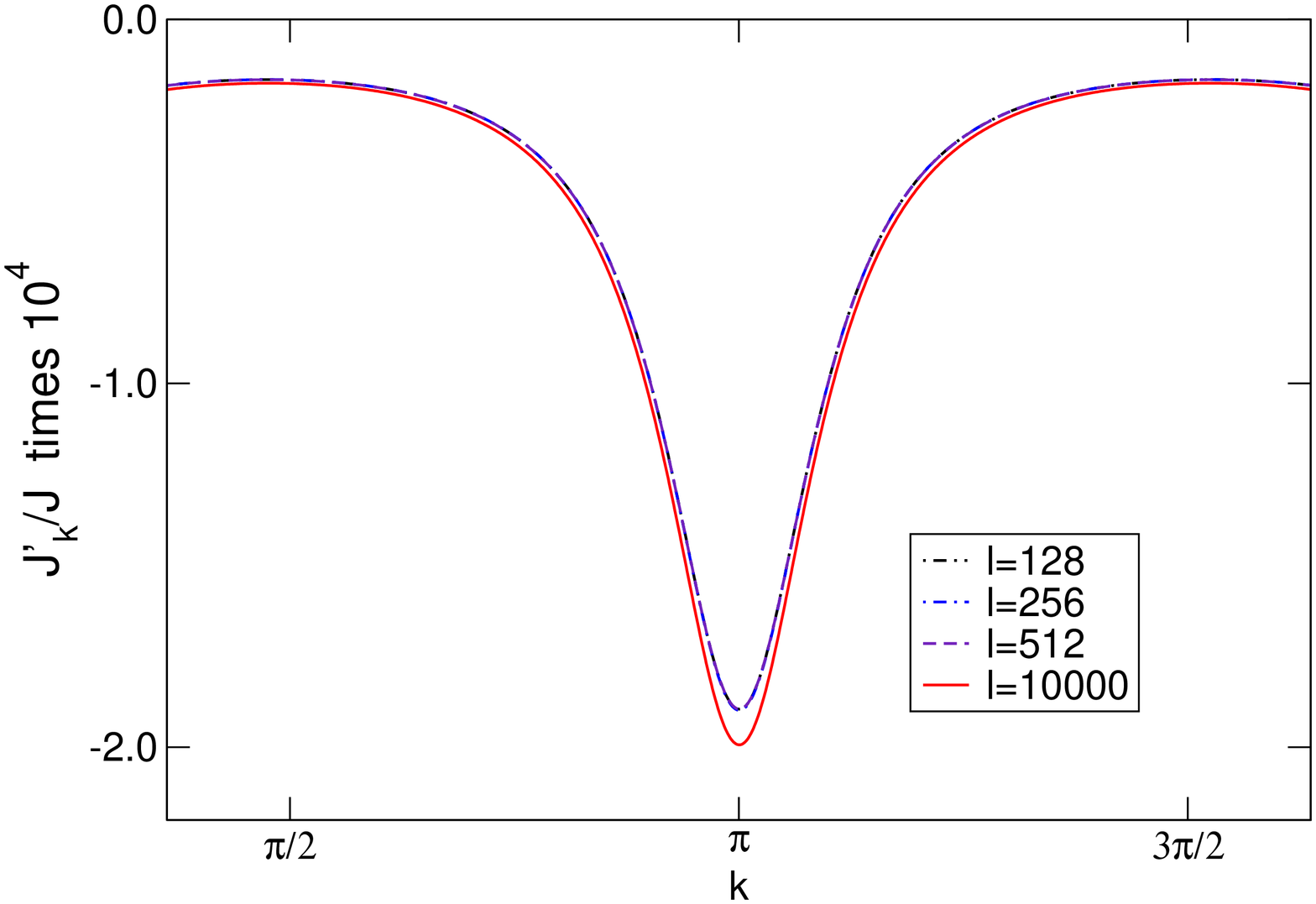}
\end{center}
\caption{Kondo coupling $J'_k$ as a function of the momentum $k$ in
the vicinity of $k=\pi$ for several values of the parameter $l$
characterizing the flow. The initial value is $J'_k=0.05J$. The
couplings $J'_k$ tend to a small but finite limit for large values of
$l$. }
\label{Jp-ksmall}
\end{figure}

\begin{figure}[ht]
\begin{center}
\noindent
\epsfxsize=0.45\textwidth
\epsfbox{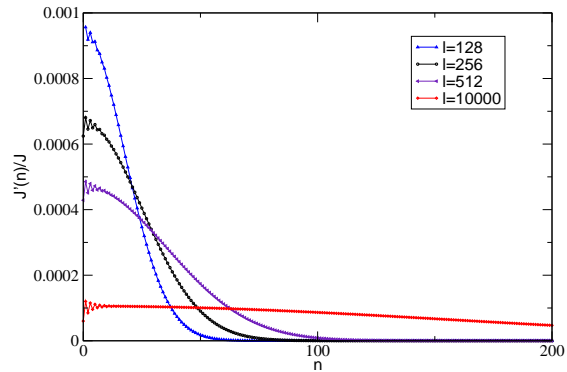}
\end{center}
\caption{Kondo coupling $J'(n)$ in coordinate space for valious values
  of the flow parameter $l$. The initial value of the Kondo coupling
  is  $J'=0.05J$ . There is a long range smooth component as well as a
  shorter range staggered one.}
\label{kondox}
\end{figure}

\begin{figure}[ht]
\begin{center}
\noindent
\epsfxsize=0.45\textwidth
\epsfbox{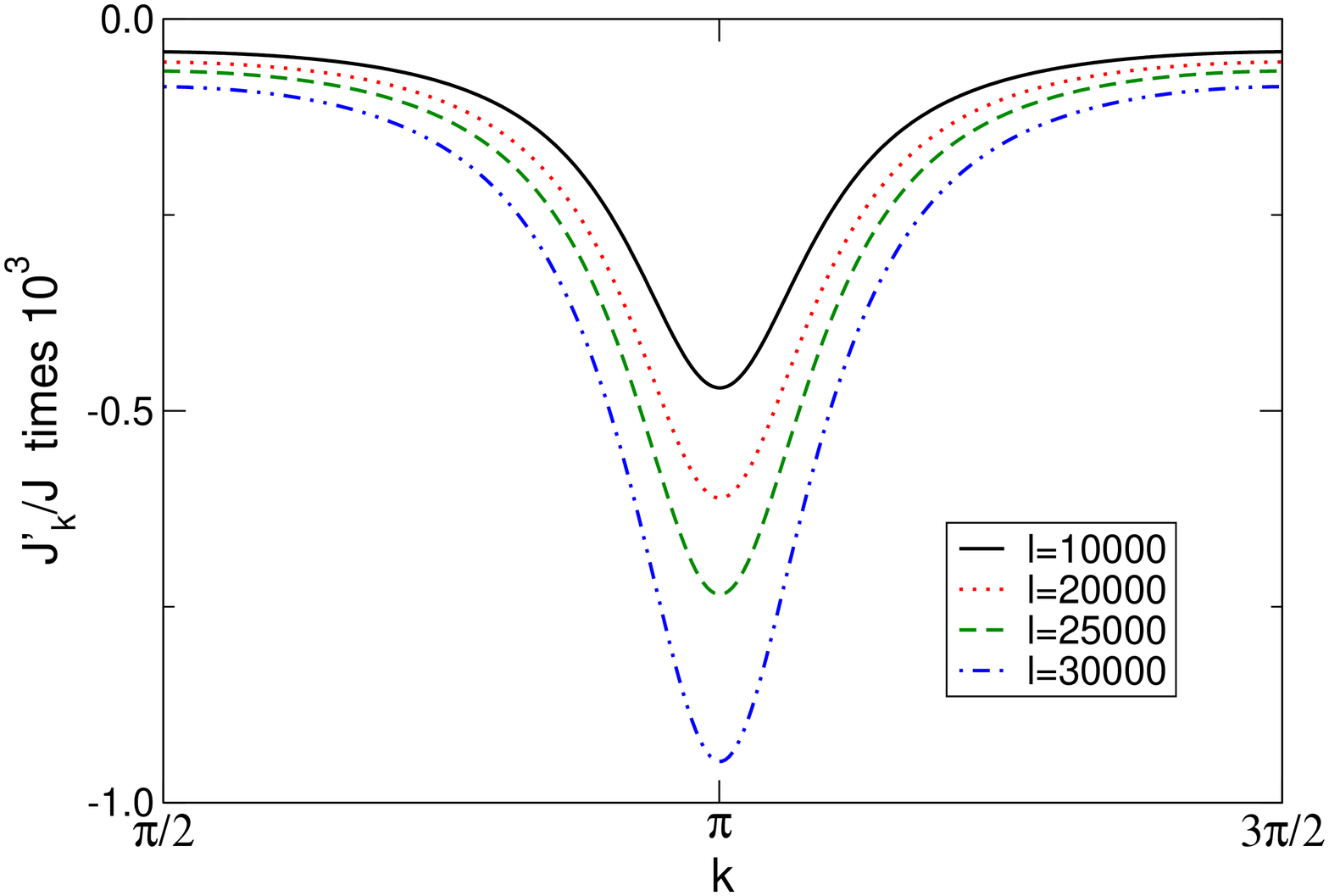}
\end{center}
\caption{
Kondo coupling $J'_k$ as a function of the momentum $k$ in
the vicinity of $k=\pi$ for large values of the parameter $l$
characterizing the flow. The initial value is $J'=0.06J$. The
absolute values of the couplings are seen to increase with $l$.
}
\label{runaway}
\end{figure}

We find that for $l\agt 100$ and small initial values of the Kondo
interaction $|J'|<0.1J$, the couplings $J'_k$ become very small
everywhere except at $k= 0,2\pi$. This holds in the ferromagnetic as
well as the antiferromagnetic case. This behaviour is shown in
Fig.\ref{Jp-k}, where we plot $J'_k$ as a function of $k$ for
different values of $l$ and $J'=0.05J$. The vicinity of $k=2\pi$ is
shown in greater detail in Fig.\ref{Jp-k2pi}. 
In Fig.\ref{Jp-ksmall} we show the Kondo coupling in the vicinity of
$k=\pi$ for several values of the flow parameter $l$. We observe that
for large values of $l$ $J'_k(l)$ appears to approach a small but
finite limit. 
In position space the Kondo coupling $J'$ becomes more and more
long-ranged as the flow parameter $l$ increases. This is shown in
Fig.\ref{kondox}. 
For very large values of the flow parameter $l$ we always enter a
regime in which the $J'_k$ diverge. This is shown for the case
$J'=0.06J$ in Fig.\ref{runaway}.

\subsubsection{RKKY Interaction}
\begin{figure}[ht]
\begin{center}
\noindent
\epsfxsize=0.45\textwidth
\epsfbox{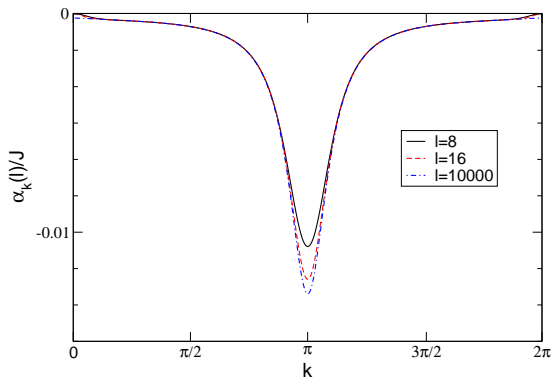}
\end{center}
\caption{RKKY coupling $\alpha_k$ as a function of the momentum $k$
for several values of the parameter $l$ characterizing the flow. The
initial value is $J'_k=0.05J$.} 
\label{alphak}
\end{figure}

The flow of the RKKY couplings $\alpha_k(l)$ is shown in
Fig.\ref{alphak} for several values of the flow parameter $l$. We see
that the flow appears to approach a small finite limit quite
quickly. The RKKY couplings are strongest in the vicinity of $k=\pi$,
where they become of order $J'^2/J$.

For sufficiently large values of the flow parameter $l$ the RKKY
interaction starts to decrease quite quickly and eventually diverges.

\subsubsection{Heisenberg Interaction}
\begin{figure}[ht]
\begin{center}
\noindent
\epsfxsize=0.45\textwidth
\epsfbox{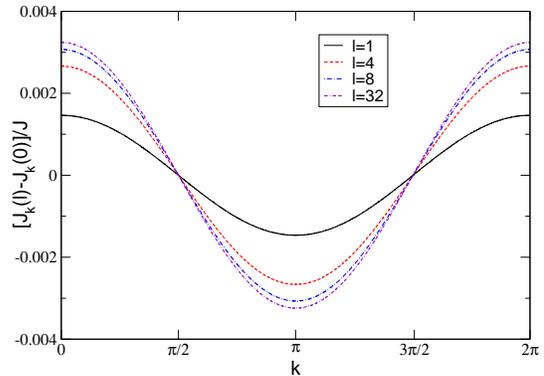}
\end{center}
\caption{Difference of Heisenberg couplings $J_k(l)-J_k(0)$ as a
function of the momentum $k$ for several values of the parameter $l$
characterizing the flow $l=1$, $l=4$, $l=8$, $l=32$.
The initial value was taken to be $J'_k=0.05J$.}
\label{deltaJp}
\end{figure}

The Heisenberg interaction changes only weakly under the flow. We
therefore plot the difference $J_k(l)-J_k(0)$ rather than $J_k(l)$
itself in Fig.\ref{deltaJp}.
We find that the $k$ dependence is unchanged under the flow, so that
\be
J_k(l)=J(l)\cos(k)\ .
\ee
Here $J(l)=J+{\cal O}\bigl(J'^2/J\bigr)$.

\section{Discussion}
\begin{figure}[ht]
\centering
\epsfxsize=0.45\textwidth
\epsfbox{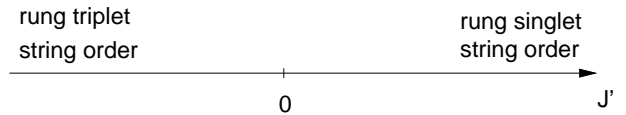}
\caption{Zero temperature phase diagram of the SU(2) invariant
one dimensional Kondo necklace model.}
\label{fig:phase}
\end{figure}

The above analysis of the flow equations shows that a small Kondo
interaction drives the system to a strong coupling regime. 
On the basis of the analysis presented in this work we cannot
establish the nature of the strong coupling phase. However, using
exact diagonalization of small systems one can see that there is a
spin gap for $J'\agt 0.3 J$, which suggests that the large-$J'$ phase
extends at least down to such interaction strengths. Combining this
observation with the result of the flow equation analysis suggests
a form of the phase diagram as shown in Fig.\ref{fig:phase}. 
This would imply there presence of a spin gap $\Delta$ for any $J'\neq
0$, but the flow equation approach presented here does not allow for a
determination of how $\Delta$ scales with $J'$.

\begin{acknowledgments}
We thank F. Assaad, M. Kiselev and D. Schuricht for
important discussions and F. Assaad for providing us with results
prior to publication. 
This work was supported by the EPSRC under grant EP/D050952/1 (FHLE
and TK) and by the US DOE under contract \#DE-AC02-98CH10886 (IZ).

\end{acknowledgments}

\appendix

\section{Commutators}
\label{app:comm}
In this appendix we list the commutators needed for determining the
flow equations.
\bea
[\eta_1,H_0]&=&\frac{1}{N^3}\sum_{\vk,\vp,\vq}\Gamma^{(1)}_{k,p,q}\
{\bf S}_{\vk+\vp}\times{\bf S}_{-\vk}\cdot
{\boldsymbol\sigma}_{\vq-\vp}\times{\boldsymbol\sigma}_{-\vq}\nn
&+&\frac{1}{N^3}\sum_{\vk,\vp,\vq}\Gamma^{(2)}_{k,p,q}
{\bf S}_{\vq}\cdot{\bf S}_{-\vk}\
{\bf S}_{\vk+\vp-\vq}\cdot{\boldsymbol\sigma}_{-\vp}.
\eea
\bea
[\eta_2,H_0]&=&\frac{1}{N^3}\sum_{\vk,\vp,\vq}
\Gamma^{(3)}_{k,p,q}
{\boldsymbol\sigma}_{\vk+\vp}\times{\boldsymbol\sigma}_{-\vk}\cdot
{\bf S}_{\vq-\vp}\times{\bf S}_{-\vq}\nn
&+&\frac{1}{N^3}\sum_{\vk,\vp,\vq}
\Gamma^{(4)}_{k,p,q}
{\boldsymbol\sigma}_{\vq}\cdot{\boldsymbol\sigma}_{-\vk}\
{\boldsymbol\sigma}_{\vk+\vp-\vq}\cdot{\bf S}_{-\vp}.
\eea
\bea
[\eta_1,H_1]&=&\frac{1}{N^3}\sum_{\vk,\vp,\vq}
\Gamma^{(5)}_{k,p,q}\Big\{
{\bf S}_{\vk+\vp+\vq}\cdot{\boldsymbol\sigma}_{-\vp}\
{\bf S}_{-\vk}\cdot{\boldsymbol\sigma}_{-\vq}\nn
&&\hskip 10pt
+ {\bf S}_{\vk+\vp}\cdot{\bf S}_{\vq}\
{\bf S}_{-\vk}\cdot{\boldsymbol\sigma}_{-\vp-\vq}\nn
&&\hskip 10pt
- {\bf S}_{\vk+\vp+\vq}\cdot{\bf S}_{-\vk}\
{\boldsymbol\sigma}_{-\vq}\cdot{\boldsymbol\sigma}_{-\vp}\nn
&&\hskip 10pt
-i\left[{\bf S}_{\vk+\vp+\vq}\times{\bf S}_{-\vk}\right]
\cdot{\boldsymbol\sigma}_{-\vp-\vq}+{\rm h.c.}\Big\}.
\eea
\bea
\Gamma^{(1)}_{k,p,q}&=&J'_{\vp}\left[J_{\vk}-J_{\vk+\vp}\right]
[\alpha_{\vq}-\alpha_{\vp-\vq}],\nn
\Gamma^{(2)}_{k,p,q}&=&-4J'_{\vp}\left[J_{\vk}-J_{\vk+\vp}\right]
[J_{\vq}-J_{\vk+\vp-\vq}],\nn
\Gamma^{(3)}_{k,p,q}&=&J'_{\vp}\left[\alpha_{\vk}-\alpha_{\vk+\vp}\right]
[J_{\vq}-J_{\vp-\vq}],\nn
\Gamma^{(4)}_{k,p,q}&=&-4J'_{\vp}\left[\alpha_{\vk}-\alpha_{\vk+\vp}\right]
[J_{\vq}-J_{\vk+\vp-\vq}],\nn
\Gamma^{(5)}_{k,p,q}&=&-J'_{\vq}J'_{\vp}\left[J_{\vk}-J_{\vk+\vp}\right],\nn
\Gamma^{(6)}_{k,p,q}&=&-J'_{\vq}J'_{\vp}
\left[\alpha_{\vk}-\alpha_{\vk+\vp}\right].
\eea
The commutator $[\eta_2,H_1]$ is obtained from $[\eta_1,H_1]$
by replacing $\Gamma^{(5)}$ by $\Gamma^{(6)}$ and interchanging
$S^\alpha\leftrightarrow\sigma^\alpha$.

\section{Bosonization}
\label{app:boson}
In order to bosonize the Hamiltonian (\ref{Hkl}) it is convenient
to tranform the spin operators to coordinate space. At weak coupling
the generated interactions are short-ranged. In a continuum
description the most relevant (in the renormalization group sense)
contributions can then be obtained by considering nearest-neighbour
terms. Interactions between next nearest neighbours will merely lead
to a renormalization of the couplings as well as generate (less
relevant) derivative terms. Using (\ref{F-tr}) and keeping only the
terms corresponding to nearest-neighbour interactions we find
\begin{widetext}
\bea
H(l)&\sim & J(l)\sum_i{\bf S}(i)\cdot {\bf S}(i+1)+
\alpha(l)\sum_i{\boldsymbol{\sigma}}(i)\cdot
{\boldsymbol{\sigma}}(i+1)\nn
&+&J'(l)\sum_i{\bf S}(i)\cdot {\boldsymbol{\sigma}}(i)
+
\widetilde{J}'(l)\sum_i{\bf S}(i)\cdot ({\boldsymbol{\sigma}}(i+1)+
{\boldsymbol{\sigma}}(i-1))\nn
&+&M^{(1)}(l)\sum_i{\bf S}(i)\cdot {\boldsymbol{\sigma}}(i+1)
\ {\bf S}(i+1)\cdot {\boldsymbol{\sigma}}(i)
+M^{(2)}(l)\sum_i{\bf S}(i)\cdot {\bf S}(i+1)
\ {\bf S}(i-1)\cdot
({\boldsymbol{\sigma}}(i)
-{\boldsymbol{\sigma}}(i+1))\nn
&+&M^{(3)}(l)\sum_i{\bf S}(i)\cdot {\bf S}(i+1)
\ {\boldsymbol{\sigma}}(i)\cdot {\boldsymbol{\sigma}}(i+1)
+M^{(4)}(l)\sum_i{\boldsymbol{\sigma}}(i)\cdot
{\boldsymbol{\sigma}}(i+1)
\ {\boldsymbol{\sigma}}(i-1)\cdot ({\bf S}(i)-{\bf S}(i+1))\nn
&+&iM^{(5)}(l)\sum_i[{\bf S}(i)\times {\bf S}(i+1)]
\cdot ({\boldsymbol{\sigma}}(i)-{\boldsymbol{\sigma}}(i+1))
+iM^{(6)}(l)\sum_i[{\boldsymbol{\sigma}}(i)\times
{\boldsymbol{\sigma}}(i+1)]\cdot ({\bf S}(i)-{\bf S}(i+1)).\label{Hx}
\eea
\end{widetext}
We now bosonize the part $H_0=J(l)\sum_i{\bf S}(i)\cdot {\bf S}(i+1)$
of the Hamiltonian (\ref{Hx}) by standard methods,
see e.g., \cite{Affleck89b,GNT,Giamarchi,EK,ZF}
\bea
S^\alpha(j)&\simeq&J^\alpha(z)+\bar{J}^\alpha({\bar{z}})
+(-1)^jn^\alpha(x)\ ,
\eea
\bea
J^+(z)&=&\frac{a_0}{2\pi}e^{-i\varphi(z)}\ ,\qquad
\bar{J}^+(\bar{z})=\frac{a_0}{2\pi}e^{i\bar{\varphi}(z)}\ ,\nn
J^z(z)&=&-i\frac{a_0}{4\pi}\partial_z\varphi\ ,\qquad
\bar{J}^z(\bar{z})=-\frac{a_0}{4\pi}\partial_{\bar{z}}\bar{\varphi}\ ,\nn
{\bf n}(x)&=&c\sqrt{a_0}
\Bigl(\cos\bigl(\frac{\Theta}{2}\bigr),-\sin\bigl(\frac{\Theta}{2}\bigr),
-\sin\bigl(\frac{\Phi}{2}\bigr)\Bigr).
\label{Sboso}
\eea
Here $z=v\tau-ix$, $\bar{z}=v\tau+ix$, $\Phi=\varphi+\bar{\varphi}$,
$\Theta=\varphi-\bar{\varphi}$ and we use a normalization such that
\be
\left\langle e^{i\alpha\varphi(z)}e^{-i\alpha\varphi(0)}\right\rangle
=z^{-2\alpha^2}\ .
\ee
The bosonized form of $H_0$ is
\be
H_0=\frac{v_s(l)}{16\pi}\int dx [(\partial_x\Phi)^2+(\partial_x\Theta)^2].
\label{H0boso}
\ee
Using operator product expansions we then can extract the dominant
parts of the various other terms in \rf{Hx}. Denoting $x=ja_0$ we have
\begin{widetext}
\bea
{\bf S}(j)\cdot\! {\bf S}(j+1)
{\boldsymbol{\sigma}}(j)\cdot\! {\boldsymbol{\sigma}}(j+1)&\sim & 
\Bigl[c_1+(-1)^jc_2\cos\bigl(\frac{\Phi}{2}\bigr)\Bigr]
{\boldsymbol{\sigma}}(j)\cdot{\boldsymbol{\sigma}}(j+1)
+\ldots ,\label{B6}\\
{\bf S}(j)\cdot\! {\boldsymbol{\sigma}}(j+1)
\ {\bf S}(j+1)\cdot\! {\boldsymbol{\sigma}}(j)&\sim & 
\Bigl[c_1+(-1)^jc_2\cos\bigl(\frac{\Phi}{2}\bigr)\Bigr]
{\boldsymbol{\sigma}}(j)\cdot{\boldsymbol{\sigma}}(j+1) 
+c_3[{\bf J}(x)-{\bf\bar{J}}(x)]\cdot{\boldsymbol\sigma}(j)+\ldots ,
\label{B7}
\\
\bigl({\bf S}(j)\times{\bf
  S}(j+1)\bigr)\cdot{\boldsymbol\sigma}(j)&\sim & 2c_3\
[{\bf J}(x)-{\bf\bar{J}}(x)]\cdot{\boldsymbol\sigma}(j)+\ldots ,
\label{B8}
\eea
\bea
{\bf S}(j)\cdot\! {\bf S}(j+1){\bf S}(j-1)\cdot
{\boldsymbol{\sigma}}(j)-
{\bf S}(j-1)\cdot\! {\bf S}(j){\bf S}(j-2)\cdot
{\boldsymbol{\sigma}}(j)\propto (-1)^j{\bf
n}(x)\cdot{\boldsymbol\sigma}(j).\label{B9}
\eea
\end{widetext}
As expected, the interactions generated under the flow are simply all terms
that are compatible with the global spin rotational $SU(2)$
symmetry. The term $({\bf J}(x)-\bar{\bf J}(x))\cdot
\boldsymbol{\sigma}(j)$ breaks the left-right (chiral) symmetry of the
free boson Hamiltonian in the $J'\to 0$ limit. However, it is well
known \cite{Affleck89} that interchange of left and right moving
bosons is not a symmetry of the spin-1/2 Heisenberg model at low
energies due to the presence of a marginally irrelevant interaction of
spin currents (which we have omitted in \rf{H0boso} for the sake of
brevity). Inspection of the scaling dimensions of the bosonic parts of the
various terms generated along the flow suggests that at low energies
the most relevant interaction of staggered magnetizations is
precisely the one picked out by our decoupling scheme (\ref{fluct}).
We note that the term \rf{B8} is not included in our
decoupling scheme. While it is likely to be as relevant as
${\boldsymbol{\cal J}}\cdot \boldsymbol{\sigma}$, the bare coupling
of the latter is much larger, which justifies neglecting the former in
the weak coupling regime $J'\ll J$.

Another fluctuation induced contribution that is not included in
our decoupling scheme is the interation of dimerizations
\be
(-1)^j\cos\bigl(\frac{\Phi(x)}{2}\bigr)\
{\boldsymbol{\sigma}}(j)\cdot{\boldsymbol{\sigma}}(j+1).
\label{dimer}
\ee
It is less relevant than the terms we keep. If it were the dominant
interaction in the problem, its effect would be to open up
dimerization gaps among the Kondo spins and spin-chain spins respectively.

\end{document}